\documentclass[a4paper, amsfonts, amssymb, amsmath, reprint, showkeys, superscriptaddress, nofootinbib, twoside, floatfix, aps, pra]{revtex4-1}
\usepackage[english]{babel}
\usepackage[utf8]{inputenc}
\usepackage[colorinlistoftodos, color=green!40, prependcaption]{todonotes}
\usepackage{amsthm}
\usepackage{mathtools}
\usepackage{blkarray}
\usepackage{amsmath}
\usepackage{physics}
\usepackage{xcolor}
\usepackage{graphicx}
\usepackage{subcaption}
\usepackage[left=23mm,right=13mm,top=35mm,columnsep=15pt]{geometry} 
\usepackage{adjustbox}
\usepackage{placeins}
\usepackage[T1]{fontenc}
\usepackage{lipsum}
\usepackage{csquotes}
\usepackage{tikz}
\usetikzlibrary{decorations.pathreplacing,calc}
\usepackage{qcircuit}
\usepackage[unicode=true, colorlinks=true]{hyperref} 
\usepackage{multirow}
\usepackage{ragged2e}
\begin{document}
\title{Improved Grid-Based Simulation of Coulombic Dynamics}

\author{Xiaoning Feng}
    \email{xiaoning.feng@chem.ox.ac.uk}
    \affiliation{Department of Chemistry, University of Oxford, Oxford OX1 3TA, United Kingdom}
\author{Hans Hon Sang Chan}
    \affiliation{Department of Materials, University of Oxford, Oxford OX1 3PH, United Kingdom}
\author{David P. Tew}
    \email{david.tew@chem.ox.ac.uk}
    \affiliation{Department of Chemistry, University of Oxford, Oxford OX1 3TA, United Kingdom}
    
\date{\today} 

\begin{abstract}
Accurate time-dependent quantum dynamics of Coulombic systems on grid-based representations remains computationally demanding due to the singularity of the Coulomb potential, which necessitates extremely fine spatial grids to mitigate discretisation errors. We propose two complementary correction schemes that, under identical resource budgets, consistently outperform the uncorrected counterparts. The first scheme modifies the potential operator to incorporate grid-basis structure into its representation, while the second introduces a corrected initial wavefunction inspired by analytical solutions of softened Coulomb potentials. Applied to hydrogenic systems, these corrections deliver improved energy accuracy and time fidelity across long evolutions. Beyond classical simulations, the proposed framework aligns naturally with quantum computing architectures, where the corrected operators and states can be encoded through truncated Walsh and Fourier series expansions. A resource analysis for the representative 2D hydrogen system yields a circuit depth of $1.5\times10^{8}$ gates over 6,000 Trotter steps. This study thus establishes practical strategies toward high-accuracy Coulombic dynamics on both classical and emerging quantum platforms.

\end{abstract}

\maketitle

\section{Introduction}


Studying the quantum behaviour of Coulombic systems through direct simulation of the full wavefunction dynamics without invoking the Born--Oppenheimer approximation is at once the most conceptually simple approach, and yet extremely challenging to realise. While such simulations can provide detailed insight into electron-molecule scattering, absorption cross-sections, Auger decay and other fundamental processes in light-matter interaction and chemical reactivity, the complexity of the simulation grows exponentially with the number of particles and with the fidelity of the wavefunction representation, which presents severe challenges to the application of the approach.

A wide range of simulation techniques for Coulombic systems on classical computers have been developed and studied in depth over the years~\cite{Gan2025,Schrader2024,Nys2024,Niklas2021,Rowan2020,Schutt2019,molnar2002}. Researchers have sought to design reduced-cost strategies that combine compact wavefunction representations with practical time evolution protocols. Grid‐based approaches~\cite{Schafer2024,Li2024,Sereda2024,Silaev2023,Peng2006,Gordon2006,kawata1999}, such as finite difference or discrete variable representations, directly discretise the spatial domain, but limitations in the grid density introduce sizeable errors arising from poor resolution near the singularities of the electron-nucleus Coulomb interaction. Spectral methods that employ tailored orthogonal polynomial bases~\cite{Dong2022,King2018,Zhao2023} can more accurately capture bound state features; however, they introduce artificial boundary conditions that distort continuum states. Path Integral techniques~\cite{Dornheim2023,Schoof2011,Bonitz2003} use imaginary time propagation to handle singular potentials, but this comes at the cost of resolving real-time dynamics. Quantum Trajectory theory~\cite{Svensson2023,Oriols2007,Popruzh2008} evolves wavepackets in time via stochastic averaging, but fails to resolve multi-body quantum correlations.

The exponential growth of Hilbert space complexity with system size forces a compromise in accuracy when using classical approaches. Quantum computing offers a paradigm shift for overcoming this limitation. This advantage stems from a quantum computer's capacity to encode an exponentially large Hilbert space using a linear number of qubits, and to leverage the inherent quantum superposition, entanglement and interference in the simulated dynamics~\cite{Brown2010,Bauer2023,Temme2011}. Specifically, for grid-based methods, the computational basis states of a quantum computer encode wavefunctions in discrete real space as:
\begin{equation}
\begin{split}
\underbrace{\sum^{2^n-1}_{j=0} c_j\ket{x_j}}_\text{pixels in grid-based space} = \underbrace{\sum^{2^n-1}_{j=0} c_j\ket{j}}_\text{amplitudes in quantum computer},
\end{split}
\end{equation}
with the number of stored amplitudes growing exponentially with the total qubit count $n$. The intrinsic interference between quantum amplitudes enables quantum computers to achieve exponential speed-ups over classical counterparts in simulating quantum dynamics~\cite{Childs2010,Seth1996}.

Among various numerical methods for time-evolution, the grid-based Split Operator Fourier Transform (SO-FT)~\cite{MOCKEN2008,Hermann1988,Kosloff1988} stands out for its simplicity and its ability to exploit the diagonal dominance of kinetic and potential operators in momentum and real space, respectively. By decomposing the time evolution into alternating kinetic and potential updates, SO-FT achieves impressive performance across many scenarios. It exploits the Fourier Transform to switch between complementary spaces, leading to high efficiency when targeting time-dependent dynamics or spectroscopic properties at modest scales. 

The SO-QFT approach represents a natural evolution of SO-FT towards the quantum platforms~\cite{Chan2023,Childs2022,Kivlichan2020,Kassal2008,Zalka1998}, with the use of transformative Quantum Fourier Transform (QFT)~\cite{Shor1994,Cleve1998,coppersmith2002,nielsen2010}. At its core, the SO-QFT technique retains the well-established procedure of operator splitting to separately treat kinetic and potential contributions, allowing for sequential application in their respective spaces. Consequently, this methodology benefits from mitigating the numerical difficulties associated with the non-locality of kinetic operators in the real space. By replacing the classical Fourier transform component by the exponentially faster QFT, the SO-QFT framework stands as a promising candidate for future applications on quantum computers. 

Notwithstanding the exponential compression of a $N^d$ direct-product grid of $N$-points each of $d$ dimensions onto $d \log_2 N$ qubits, large values of $N$ are still required to simulate Coulombic systems due to the necessity to properly resolve the Coulomb singularities. Insufficient resolution introduces discretisation artifacts into the simulated dynamics. 
A variety of strategies have been developed to mitigate the numerical difficulties of representing Coulombic singularities in grid-based simulations, including soft-core Coulomb potentials~\cite{TRUONG2022,Li2021,INARREA2019}, pseudopotentials~\cite{Berry_2024,Su2021,Tommi2023}, polynomial-based discrete variable representation (DVR) methods~\cite{Dong2022,King2018,Zhao2023}, and adaptive grid techniques~\cite{Boschitsch2011,Jalba2007,Lu2001}. While these approaches have proven effective in many contexts, they all face some inherent limitations. Soft-core potentials introduce empirical smoothing parameters that distort the true interaction. Pseudopotentials, polynomial DVR and adaptive grid techniques all produce highly non-diagonal potential operators that are very expensive to implement on quantum hardware. The latter two are also structurally incompatible with the efficient QFT. These drawbacks motivate more robust schemes that can preserve the underlying physics, maintain the diagonal structure of operators, and enable efficient mapping to quantum computing architectures.

In this paper we examine the discretisation error for simple model systems and propose two strategies for suppressing discretisation artifacts. These efforts are designed to address the systematic inaccuracies caused by oversimplifying Coulomb-related terms in grid-based representations. Our study not only improves classical simulation fidelity, but also proposes gate-efficient mappings of the correction schemes essential for adaptation to fault-tolerant quantum hardware.

This paper is organized as follows. Section~\ref{FoundFrame} details the foundational framework underlying our approach. Section~\ref{CorreSch} introduces the correction schemes employed to better accommodate Coulombic dynamics. Section~\ref{ClassRes} presents classical emulation results to illustrate the efficacy of these corrections. Section~\ref{QuanMap} outlines the corresponding quantum circuit architectures, along with an analysis of their quantum resource demands.

\section{Foundational Framework}
\label{FoundFrame}

We begin by detailing the basic theoretical framework of our approach. This includes the spatial-grid scheme for quantum state representation, the systems studied in this work, the SO-QFT method for time evolution, and the extraction of time-dependent properties.

\subsection{Grid-Based State Representation}

In both classical and quantum computational frameworks, grid-based representations frequently utilise plane wave basis functions due to their completeness, orthogonality, and compatibility with periodic boundary conditions.  
Specifically, we consider a truncated momentum basis of finite $2\rho$ states, $|k\rangle$, each corresponding to a plane wave function:
$$\phi_k(x)=L^{-\frac{1}{2}}e^{i\frac{2\pi x}{L}k},$$
with $k \in {-\rho, \dotsc, \rho-1}$. Here $L$ represents the length of the position domain, which is chosen to ensure negligible boundary amplitude during the time evolution. Upon application of the inverse Fourier transform, these momentum-space basis states are mapped to a set of dual states $|j\rangle$:
$$\phi_j(x)=\frac{1}{\sqrt{2\rho}}\sum_{k=-\rho}^{\rho-1}e^{-i\frac{2\pi k }{L}x_{j}}\phi_k(x),$$
where each `pixel function' $\phi_j(x)$  in the position space is sharply peaked at the grid position $x_j = \frac{jL}{2\rho}$.
As $2\rho \to \infty$, these pixel functions converge to exact Dirac delta functions: $\phi_j(x) \to \delta(x - x_j)$. In practical use, each position grid point is associated with a single amplitude of the wavefunction, assuming that the pixel function is negligible away from its central peak (i.e., $\phi_j(x)=0$ at all $x_{m\neq j}$ points and only peaked at $x_{j}$)~\cite{Chan2023}.

\subsection{Model Systems}

Our analyses focus on two characteristic simple Coulombic systems, which are small enough to allow simulations with very dense grids and long-time evolutions with short time steps. 
The first is the 2D hydrogen system~\cite{Yang1991}, where the nucleus is treated as a fixed Coulombic center at the origin and the electron is described by a ground-state quantum wavepacket
\begin{equation}
    \Psi_\text{hydrogen}=Ce^{-2\mathbf{r}},
\end{equation}
and the 2D Hamiltonian
\begin{equation}
    H_\text{hydrogen}=\underbrace{-\frac{1}{2}\frac{\partial^2}{\partial x^2}
    -\frac{1}{2}\frac{\partial^2}{\partial y^2}\ }_{K}
    +\underbrace{\frac{1}{\mathbf{r}}}_{V}\ ,
\end{equation}
leading to to an analytical energy of -2 a.u.
Here $\mathbf{r}=\sqrt{x^2+y^2}$ and $C$ incorporates all constant factors in the wavefunction.
The simulation box, discretised along two spatial dimensions (each 10~a.u. in length), determines the resolution of the electron’s position, with higher grid density yielding finer spatial details.  

The second system involves two interacting electrons confined on a one-dimensional ring with the radius of $R = 1/2$ and the analytical energy of 2.25~a.u.~\cite{Loos2012}. Here, both the wavefunction,
\begin{equation}
    \Psi_\text{2e}=\mathbf{u}(1+\frac{\mathbf{u}}{2R})^{1/2},
\end{equation}
and the Hamiltonian
\begin{equation}
    H_{2e}=\underbrace{-\frac{1}{R^2}\frac{\partial^2}{\partial \omega^2}\ }_{K} + \underbrace{\frac{1}{\mathbf{u}}}_{V}\ ,
\end{equation}
with $\mathbf{u}=R\sqrt{2-2\cos{\omega}}$,
depend on the angular separation $\omega$ between electrons, which is defined as the intracule angle coordinate spanning $[0, 2\pi]$. This angular variable is discretised to control the spatial resolution. Atomic units are used for all simulations.

Discretisation errors occurs when these spatially continuous potential and wavefunction are represented by their values on discrete grid points. The singularity of either attractive or repulsive Coulomb potential also poses numerical difficulties. To prevent divergence, the grid is defined to avoid including this point explicitly. This exclusion, along with the sparse nature of grid-based discretisation, introduces artifacts in ground-state energies and time evolution fidelity. Denser grids mitigate this issue but at the expense of increased computational resources.

\subsection{Time Propagation}

In the SO-QFT method, the time evolution is decomposed into alternating kinetic and potential operator actions via the symmetric Trotter-Suzuki formula, from which errors occur due to the non-commutativity between partitioned Hamiltonian terms. Adopting the second order form~\cite{Suzuki1976,suzuki1985,Hatano2005} bounds this error scaling within $\mathcal{O}(dt^{3})$:
\begin{equation}
e^{-iHdt} = e^{-iVdt/2}e^{-iKdt}e^{-iVdt/2}+\mathcal{O}(dt^{3}),
\end{equation}
where $dt$ denotes the discretisation interval between adjacent time steps. This symmetric workflow proceeds through (i) a half-step potential operator $e^{-iVdt/2}$ in the position space, (ii) a full-step kinetic operator $e^{-iKdt}$ executed via QFT-mediated momentum-space propagation, and (iii) a concluding half-step potential update $e^{-iVdt/2}$. All simulations in this work are conducted under a sufficiently high time resolution, e.g., 30,000 steps ($dt=0.0002$~a.u.) for the 2-electron quantum ring and 6,000 steps ($dt=0.001$~a.u.) for 2D hydrogen over a total duration of 6~a.u. The total simulation time is chosen to be long enough to enable extraction of $E_\text{dynamic}$ and assessment of fidelity over physically meaningful timescales. The time step sizes are selected to minimise the impact of time discretisation and enable a focused evaluation of spatial resolution effects.

Conversions between position and momentum spaces are typically performed using Fourier Transform. On classical platforms, this process is commonly realised via the fast Fourier Transform (FFT), which scales as $\mathcal{O}(n2^n)$ for a vector of size $N = 2^n$. In contrast, the QFT provides an exponentially more efficient mechanism, requiring only $\mathcal{O}(n^2)$ gates for $n$-qubit systems. 


\subsection{Property Extraction}

Within the SO-QFT framework, the total evolution time is discretised into finite steps. We monitor two key properties during time evolution: the grid-dependent eigenenergy obtained from dynamical simulations and the evolution fidelity. Both are derived from the time-dependent autocorrelation function, 
\begin{equation}
A(t)=\langle\Psi(t)|\Psi(0)\rangle=e^{iE_\text{dynamic}t}, 
\end{equation}
which is the inner product of the time-evolved state at time $t$ with the initial state at $t = 0$~\cite{Nauen1990}.
The cosine periodicity of its real component, $\cos{(E_\text{dynamic}t)}$, enables retrieval of this numerical energy, $E_\text{dynamic}$, which corresponds to the eigenenergy of the discrete grid Hamiltonian and serves as a robust indicator for algorithmic accuracy. 
When the initial state is precisely the eigenstate of the grid Hamiltonian, $E_\text{dynamic}$ matches the static expectation value $\bra{\Psi}H\ket{\Psi}$. For an exact simulation with perfect representation of the Hamiltonian and wavefunction, $E_\text{dynamic}$ is expected to be the analytical ground-state energy.
Fidelity, on the other hand, is evaluated by the absolute value of the autocorrelation function. For systems initiated in eigenstates of the Hamiltonian, this modulus should ideally remain unity throughout time evolution: $$\lvert A(t)\rvert =\lvert \langle\Psi_{n}(t)|\Psi_{n}(0)\rangle\rvert=1.$$ Any deviation indicates fidelity loss and thus provides a quantitative measure of temporal coherence in the simulated dynamics.

 

\section{Correction Schemes}
\label{CorreSch}


In conventional grid-based simulations, potentials are applied pointwise on a discretised spatial grid, implicitly assuming each basis function as a Dirac delta function. While a small number of simulation grid points suffice for smooth fields, e.g. quadratic potentials, the inability to resolve the singularities near their origins introduce errors when representing dynamics with the Coulomb interaction, particularly under a limited grid resolution.
The immediate assumption that a higher spatial resolution via an increased grid density leads to improved accuracy is widely held. Yet, this direct scaling of resolution, while conceptually straightforward, exhibits intensive computational burden for large systems. This trade-off becomes problematic in scenarios necessitating prolonged time propagation, rendering it less than ideal for practical applications.

To address this, we propose a correction scheme in which the Coulomb potential is re-evaluated by computing the expectation value of the original Coulomb potential within the actual pixel function $\phi_j(x)$. 
At finite grid size, each $\phi_j(x)$ deviates from an ideal Dirac delta function and instead possesses a non-zero spatial extent with oscillatory side lobes.
Specifically, we define a corrected potential
\begin{equation}
\begin{split}
V_\text{corrected}&=\langle \phi_{i}(x)|V|\phi_{k}(x)\rangle \text{ (in 1D)}, \\
&=\langle \phi_{i}(x)\phi_{j}(y)|V|\phi_{l}(y)\phi_{k}(x)\rangle \text{ (in 2D)}, 
\end{split}
\end{equation}
which accounts for the spatial extent and shape of the localized pixel functions and more accurately reflects the singular potential's influence over the local region.
The matrix elements are evaluated numerically on an auxiliary grid with $N_{\text{correction}}$ points per dimension, which is independent of the simulation grid $N_{\text{simulation}}$. This correction grid serves purely as a quadrature tool and does not modify the Hilbert space dimension of the subsequent time-dependent simulation. 

While non-zero off-diagonal elements arise from the overlap between pixel functions, the full basis-averaged potential matrix is diagonally dominant, and the diagonal terms are expected to capture the majority of the correction effect. Since off-diagonal elements cannot be effectively truncated to leading terms, their inclusion would require full matrix construction and incur substantial overhead for both classical preprocessing and quantum implementation. Motivated by pragmatic considerations of computational efficiency and numerical stability, we have limited our consideration to contributions arising solely from the diagonal elements, where $i = k$ and $j = l$. This correction can be physically interpreted as recognising that the electron is not localised at an infinitesimal point $x_j$, but is represented by a finite-width wavepacket defined by the grid basis function $\phi_j(x)$. The corrected potential therefore corresponds to the basis-averaged Coulomb interaction experienced by this spatially extended state. 


The new set of potential values, $V_\text{corrected}$, is used during simulation in place of the naive point-sampling potential. 
Although the simulation itself still runs on the original coarse grid $N_{\text{simulation}}$, using a denser auxiliary $N_{\text{correction}}$ for the numerical calculation of $V_\text{corrected}$ yields significant performance improvements in both the ground-state energy integrity throughout time-dependent dynamics and the evolution fidelity as a function of time (see Section~\ref{CorreResPot}). The correction leverages higher-resolution information in the preprocessing stage while keeping the computational complexity of the main simulation unchanged. Specifically, this means the quantum circuit depth and qubit count remain identical to the uncorrected case, in contrast to increasing $N_\text{simulation}$ which would require additional qubits and exponentially more gates. 

A useful comparison may be drawn with the augmented split-operator (ASO) approach introduced by~\cite{Chan2023}, in which a unitary augmentation is derived to repair discrepancies between the exact and split-operator propagators within a selected core subspace. While ASO can yield substantial fidelity improvements, its derivation is non-trivial and requires additional circuit execution at every Trotter step. By contrast, the proposed potential correction only involves a one-off numerical quadrature to modify the discretised potential operator, leaving the SO-QFT structure and per-step gate complexity unchanged. The two approaches therefore target different sources of errors: ASO mitigates propagator-level inaccuracies, whereas our correction reduces errors arising from the discretisation of the singular potential. 

Another challenge in grid-based simulations involving Coulomb systems lies in initial state determination, which would typically be the ground state of the system in the absence of a perturbing interaction. In low-resolution grids, the ground state of the low-resolution Hamiltonian differs from the exact eigenfunction and we must decide whether to initialize with a close approximation to the eigenstate of the Hamiltonian we evolve with, or a close approximation to the eigenstate of the analytic solution. To obtain consistent and stable dynamics that converges systematically to the analytic behaviour with increasing resolution, one should choose the former and prepare an initial eigenstate of the low-resolution Hamiltonian. The low-resolution Coulombic interaction essentially replaces the sharp singularity with a softened  Coulombic interaction. Inspired by prior analytical solutions of hydrogenic systems derived from softened Coulomb potentials~\cite{Li2021}, we propose to use 
\begin{equation}\label{eqcorre}
\begin{split}
&\Psi_\text{corrected} = C\mathbf{r}^{\frac{Z}{\alpha} - b}e^{-\alpha \mathbf{r}},
\end{split}
\end{equation} 
as a corrected wavefunction to reduce the numerical mismatch between the initial state and the grid-based Hamiltonian. Here $b = 0, -\frac{1}{2}, -1$ for 1D, 2D and 3D hydrogenic models, and $\alpha = \sqrt{-2E_\text{dynamic}}$ . In this formulation, all constant coefficients are absorbed into the prefactor $C$. 
Since $E_\text{dynamic}$ varies with the grid density employed in the numerical simulation, the parameter $\alpha $ must be uniquely calculated for each specific grid density.
Thus, by running a preliminary trial simulation at the chosen spatial resolution, one can determine $E_\text{dynamic}$, calculate the corresponding $\alpha$, and subsequently derive the corrected wavefunction $\Psi_\text{corrected}$.

Our results (see Section~\ref{CorreResWav}) on the 2D hydrogen system confirm that initializing the simulation with this corrected wavefunction yields extremely high time fidelity, validating its effectiveness. As $\Psi_\text{corrected}$ more closely approximates the true physical state in grid-based space, this correction scheme enables stable time-dependent studies without requiring prohibitively dense grid resolutions.

While demonstrated here for hydrogenic systems, the strategy of modifying initial wavefunctions holds potential for broader applications beyond the current context. For other Coulombic systems where analytical eigenfunctions of analogous softened potential approximations are available, similar corrections can be deduced. For more complex molecular systems a successful scheme may involve using corrected atomic orbital expansions as good initial states, which could potentially be precomputed for individual atomic species and combined to construct molecular wavefunctions without the need to solve the full molecular problem prior to simulation. This offers a systematic means to enhance simulation reliability while avoiding impractical grid refinements, thus broadening the applicability of grid-based quantum dynamics methods. 

An alternative strategy for computing the ground state of a discretised Hamiltonian is imaginary time propagation (ITP). In a broad sense, both ITP and the present wavefunction correction seek to recover the eigenstate of a given grid-based Hamiltonian. For the model system considered in this work, it is not necessary to use ITP. Our corrected initial state provides a compact, analytically parameterised ansatz that retains an explicit functional form. Critically, it relies solely on real-time unitary evolution to extract the ground state energy needed to calibrate the grid-dependent parameter. As digital quantum computers natively support only unitary operations, this scheme is directly transferable to the quantum setting. 
Quantum implementation of ITP instead requires either post-selected measurement protocols~\cite{Leadbeater2024, Xie2024}, hybrid variational constructions~\cite{McArdle20192, Gomes2020} , or recursive schemes with exponential depth growth and discretization approximation errors~\cite{Gluza_2026}, introducing significant overhead and error accumulation. 
That said, ITP remains indispensable for more general cases. Given that the convergence rate and total cost of ITP depends strongly on the quality of the initial guess, initializing the propagation with a corrected wavefunction could provides a better starting point for ITP.

\section{Performance of Correction Schemes}
\label{ClassRes}

\begin{figure}[!htbp]
    \centering
    \includegraphics[scale=0.15]{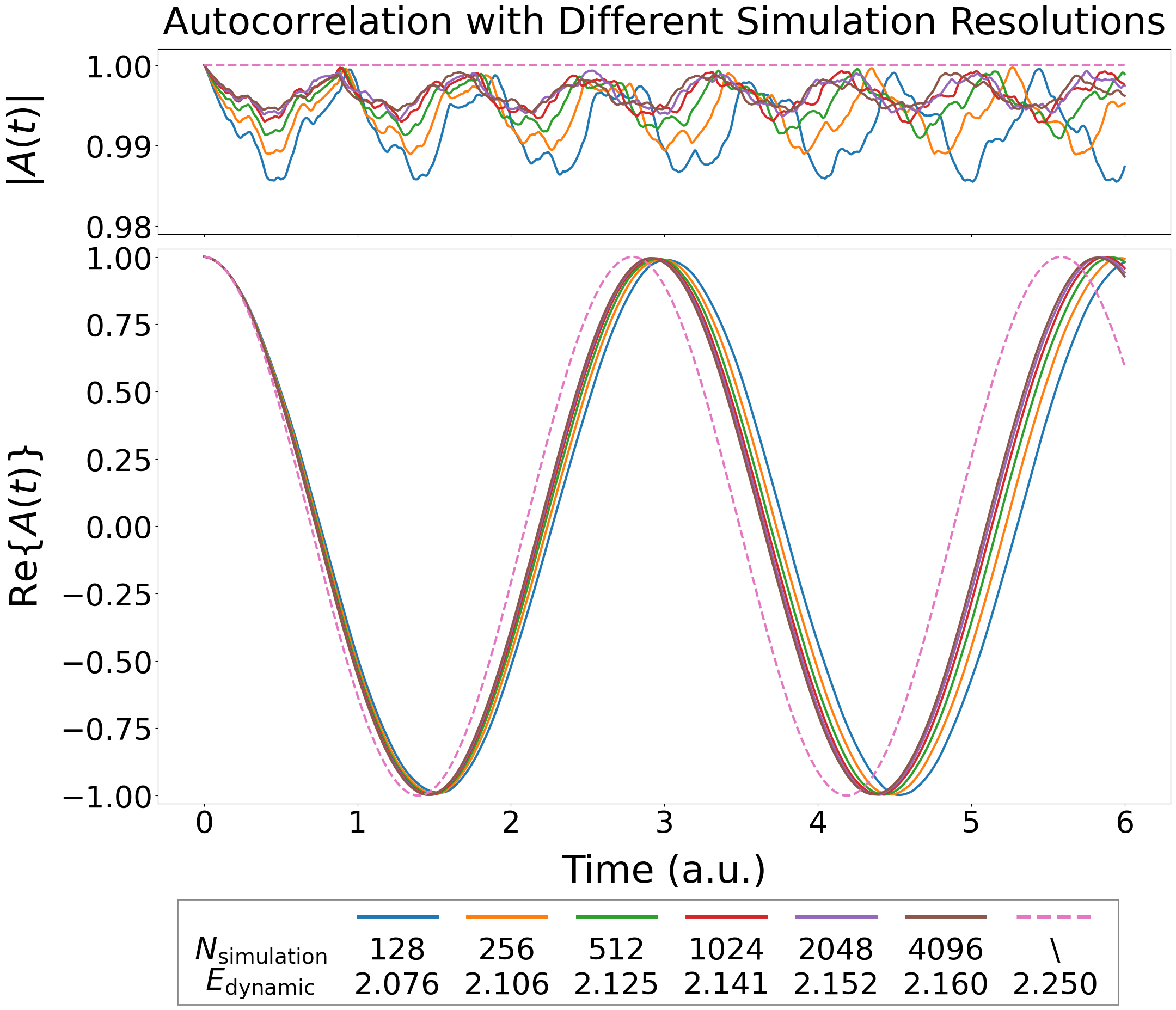}
    \caption{\justifying Absolute value and real part of autocorrelation functions recorded from simulating the 2-electron quantum ring with different $N_\text{simulation}$. Analytical profiles are plotted with dashed linestyle.
    }
    \label{ring Nsimula}
\end{figure}

We first assess the general characteristics of the SO-QFT scheme in systems featuring singular Coulombic potentials -- which we will reference as the uncorrected benchmark. As a representative case, we herein present results of the 1D two-electron quantum ring. Figure~\ref{ring Nsimula} illustrates that, in the absence of correction, both the phase difference between numerical and analytical $\operatorname{\mathbb{R}e}\{A(t)\}$ curves and the deviation of $|A(t)|$ curves from unity decrease systematically as the spatial resolution increases (i.e., the number of grid points $N_\text{simulation}$ increases within the fixed angular domain $[0, 2\pi]$).
As expected, increasing grid density improves energy estimate and time fidelity, but incurs a steep computational cost in terms of memory and gate complexity, posing a practical limitation on such a brute-force approach. This motivates the use of correction schemes that can suppress errors even for coarse grid settings.

We therefore evaluate the efficacy of our two correction schemes in improving overall simulation quality, still using two complementary indicators: the numerical ground-state energy $E_\text{dynamic}$ for algorithmic accuracy and the modulus of the autocorrelation for fidelity retention over time.

\subsection{Potential Operator Corrections}
\label{CorreResPot}

\begin{figure*}[!htbp]
    \centering
    \begin{subfigure}[b]{0.48\textwidth}
        \includegraphics[scale=0.15]{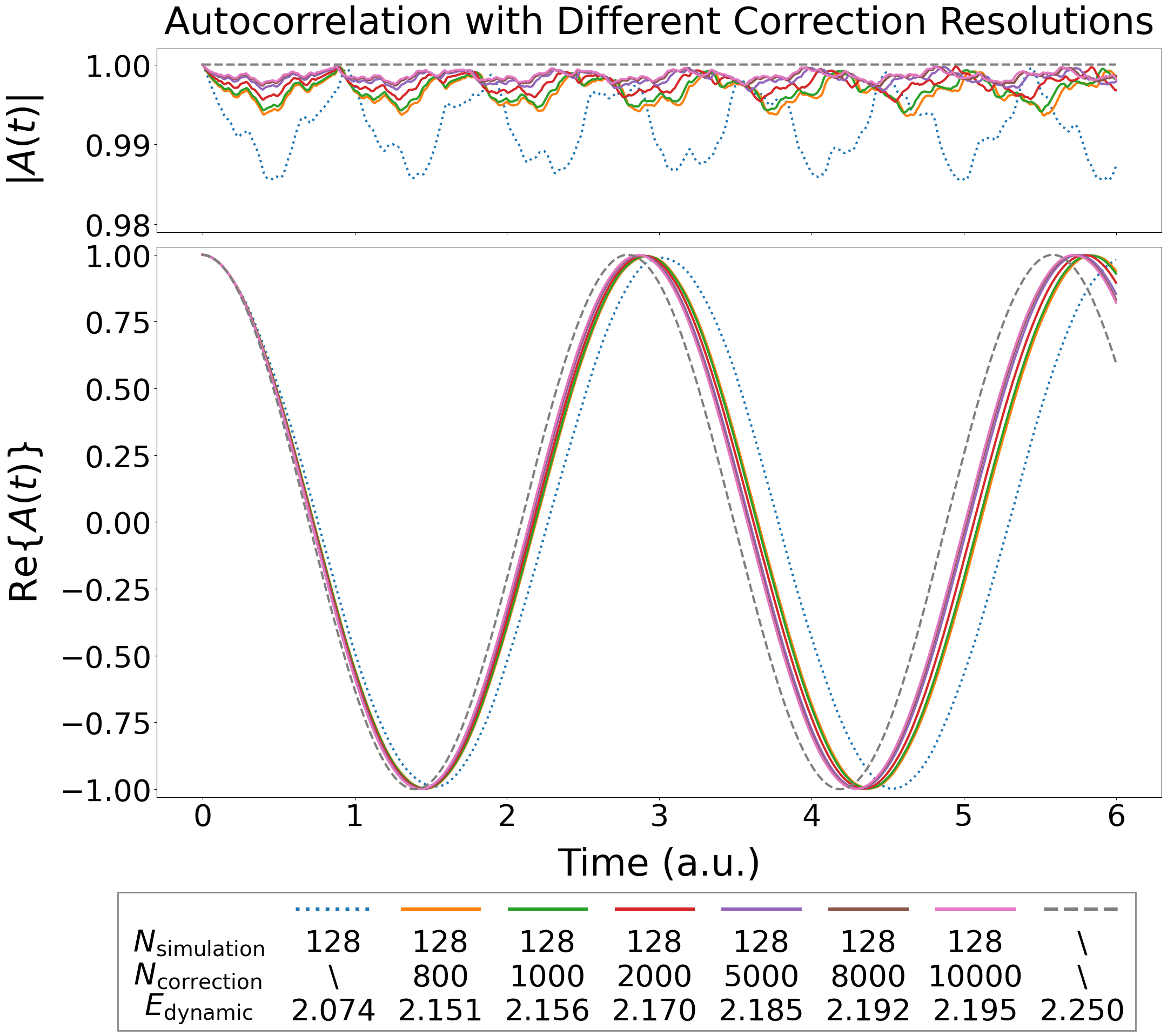}
        \caption{Autocorrelation for the 2-electron quantum ring}
        \label{ring Ncorre}
    \end{subfigure}
    \hspace{0.01\textwidth}
    \begin{subfigure}[b]{0.48\textwidth}
        \includegraphics[scale=0.15]{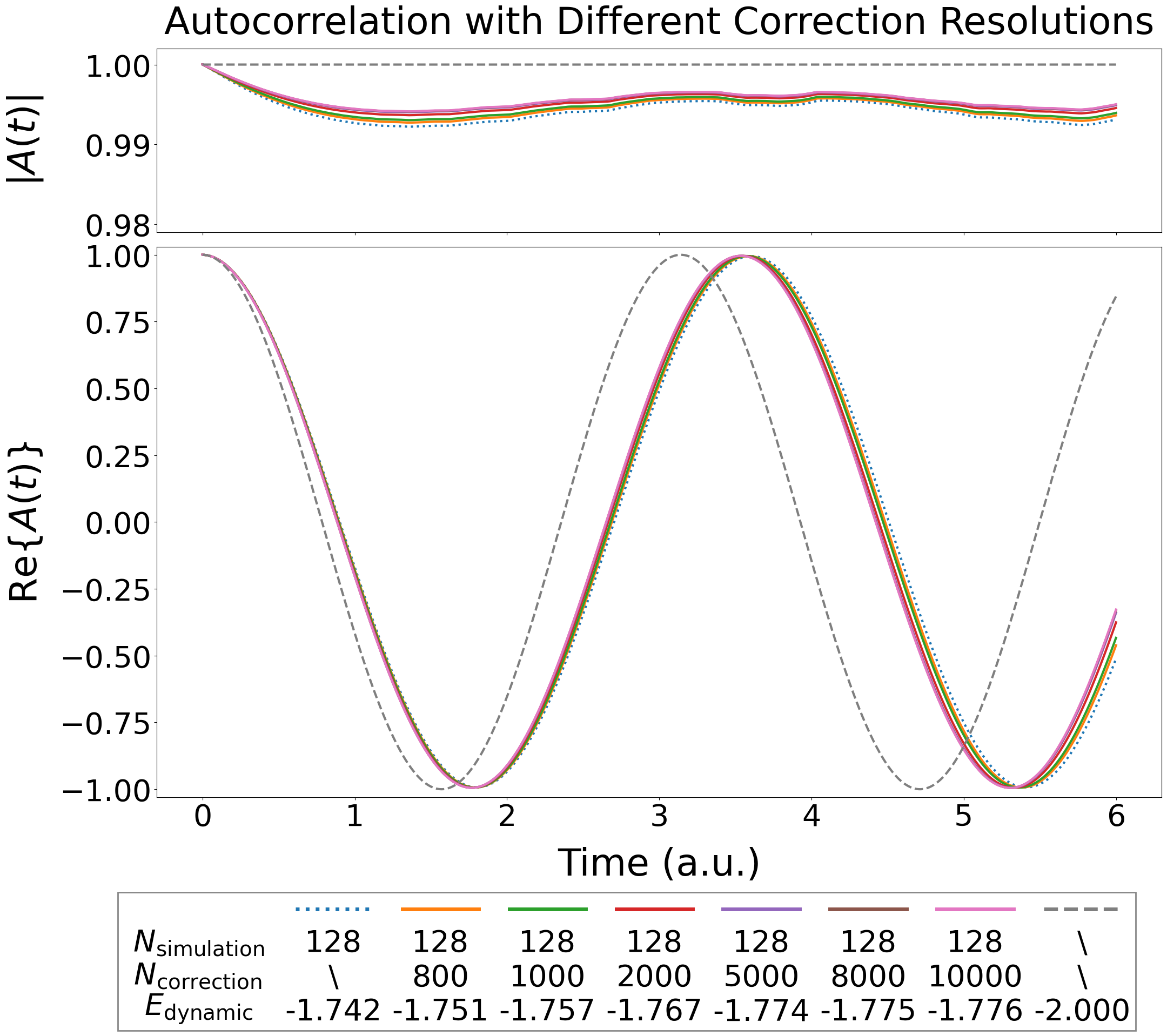}
        \caption{Autocorrelation for the 2D hydrogen}
        \label{hyd Ncorre}
    \end{subfigure}
    \caption{
        \justifying Absolute value and real part of autocorrelation functions recorded from simulations of (a) the 2-electron quantum ring and (b) the 2D hydrogen using $V_\text{corrected}$ computed from different $N_\text{correction}$. For both panels, $N_\text{simulation}$ is fixed at 128 per dimension. Analytical profiles and results from uncorrected runs are plotted with dashed and dotted linestyles, respectively.
    }
    \label{dif Ncorre}
\end{figure*}

Diagonal corrections on the potential operator decouple the grid density used for physical propagation from that used for numerical calculation of the basis-integrated $V_\text{corrected}$.
We run simulations under $V_\text{corrected}$ with the same simulation grid size $N_\text{simulation}$ per dimension (as above) but vary the number of auxiliary correction grids $N_\text{correction}$, as shown in Figure~\ref{dif Ncorre}. 

Compared to uncorrected benchmarks, these numerical results confirm that simulations using $V_\text{corrected}$ display improved agreement with analytical energies and show enhanced time fidelity for both the 1D two-electron quantum ring and the 2D hydrogen systems. This behaviour is rationalized by the fact that the corrected potential captures more of the singular feature of the Coulomb interaction that would otherwise be missed in such low-resolution simulations.

In addition, as the density of the quadrature grid $N_\text{correction}$ increases, both the extracted $E_\text{dynamic}$ and $|A(t)|$ exhibit improvements.  
By resolving more of the pixel function's structure in the correction step, we achieve better modeling of the potential landscape. 
Technically, we only need to calculate $V_\text{corrected}$ once and the computational cost involved will not be escalated during long-time propagation.


As each entry in $V_\text{corrected}$ is computed by integrating over $N_\text{correction}^d$ grid points for $d$-dimensional systems, the major limitation lies in the exponentially increasing cost with system dimensionality. 
To make this step more scalable, we explore a piecewise non-uniform grid that concentrates points near the origin (where singularities occur), and allocates coarser spacing elsewhere.

\begin{table}[h]
\centering
\begin{tabular}{|c|c|c|c|}
\hline
Central Spacing &\multicolumn{2}{|c|}{$N_\text{correction}$ within $[-5,5]$} & $E_\text{dynamic}$ \\
\hline
\multirow{2}{*}{0.002} &Non-Uniform & 1800 & -1.775 \\
\cline{2-4}
&Uniform & 5000 & -1.774 \\
\hline
\multirow{2}{*}{0.001} &Non-Uniform & 2800 & -1.777 \\
\cline{2-4}
&Uniform& 10000 & -1.776 \\
\hline 
\end{tabular}
\caption{\justifying Simulated $E_\text{dynamic}$ of the 2D hydrogen system using uniform and non-uniform grid allocations in $V_\text{corrected}$ calculation.
In the non-uniform cases, $N_\text{correction}=$ 1800 and 2800 represent that 1000 and 2000 grid points are assigned within the central region, $[-1,1]$~a.u., respectively, while the outer tails ($[1,5]$~a.u. and $[-5,-1]$~a.u.) are allocated with 800 grid points. 
}
\label{nonuniform}
\end{table}

As seen in Table~\ref{nonuniform}, the energy values simulated using the non-uniform grid are close to those using much more uniform grids, provided the central resolution is comparable. The key insight is that the dominant contributions to the potential correction arise near the singularity, justifying the use of non-uniform strategy for higher-dimensional calculations to avoid the overhead of uniformly dense grids.

This finding also rationalizes the difference in correction efficacy between 1D and 2D systems. 
Since the dominant contribution to the potential correction arises from a radially symmetric region surrounding the singularity, accuracy is determined by the quality of sampling in this small volume.
Our Cartesian grid, however, is structured along orthogonal coordinate axes, and the effective sampling density of the critical region degrades with increasing dimensionality. 
While a given grid density can adequately sample the 1D linear singular region, in 2D with the same grid spacing most radial directions within the circular critical region remain undersampled.
This geometric mismatch means that a fixed $N_\text{correction}$ per dimension provides less effective resolution in higher dimensions, explaining the poorer correction performance observed in 2D.



\subsection{Initial Wavefunction Corrections}
\label{CorreResWav}

\begin{figure}[!htbp]
    \centering
    \includegraphics[scale=0.15]{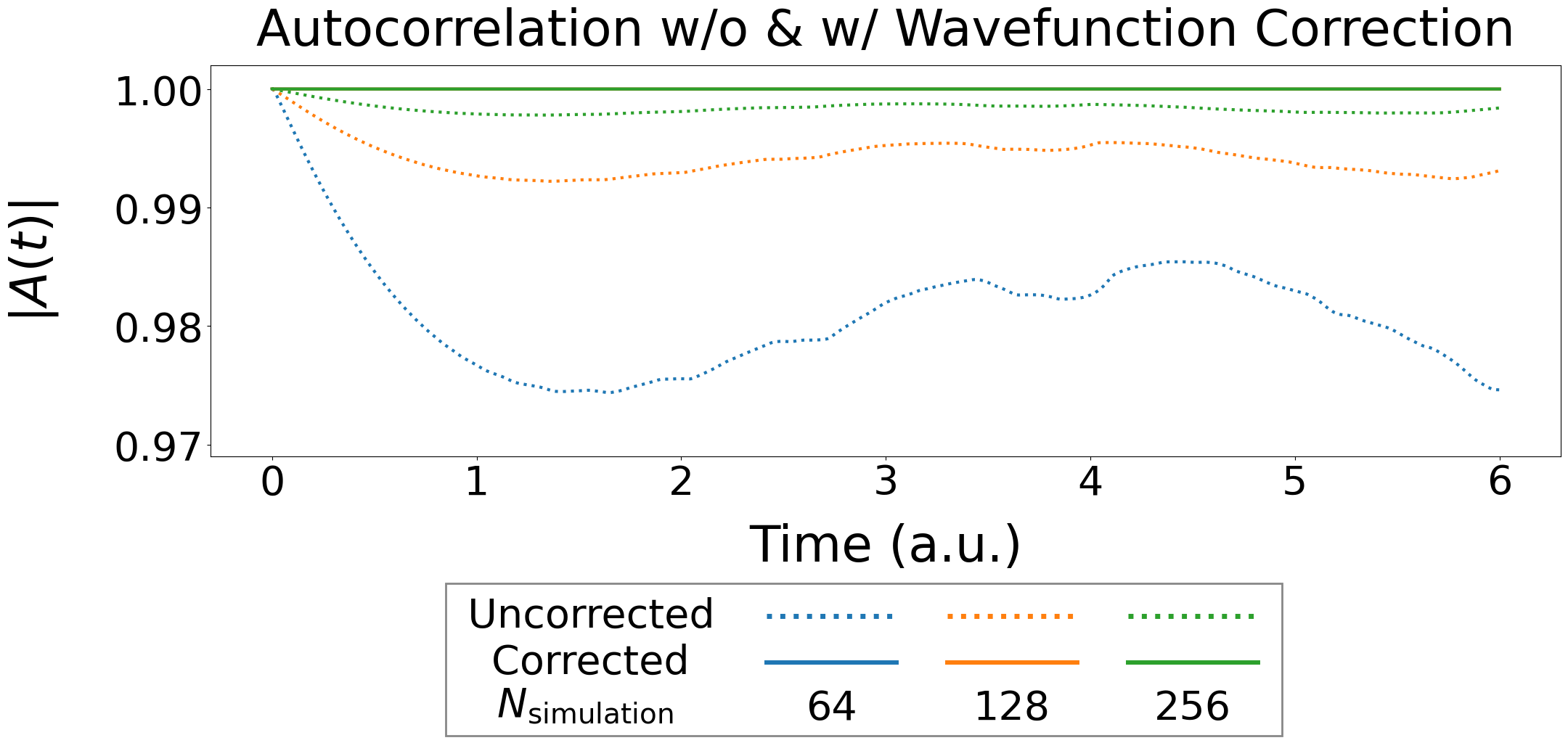}
    \caption{\justifying Absolute value of autocorrelation functions recorded from 2D hydrogen simulations initialized with uncorrected and corrected wavefunctions across different $N_\text{simulation}$.}
    \label{wavef corr}
\end{figure}

Here we evaluate the impact of wavefunction correction on time propagation fidelity.
In Figure~\ref{wavef corr}, simulations initialized with corrected wavefunctions demonstrate near-perfect time fidelity. 
Although the deviations of the uncorrected $|A(t)|$ from unity appear small within the short time window shown, they indicate a residual excited-state admixture in the initial condition. Such contamination constitutes a coherent, systematic error that persists throughout the time evolution and may accumulate over extended propagation times. The present correction provides a controlled means of eliminating this mismatch at small additional cost. 
Across spatial grids of varying resolution examined ($N_{\text{simulation}} = 64, 128, 256$), the corrected cases keep tightly aligned with the ideal $|A(t)| = 1$ benchmark, in stark contrast to the visible deviations seen in the uncorrected runs. 
This indicates that the correction effectively recovers eigenstate-like stability and that this benefit extends to various values of $N_\text{simulation}$, confirming the robustness of utilising corrected initial states for maintaining high time fidelity even on coarse grids. 


The results in Figure~\ref{wavef corr} were obtained without applying the potential correction. The performance of wavefunction correction is independent of whether the potential correction is employed, as in each case the corrected state is constructed from the corresponding $E_\text{dynamic}$ of the specified Hamiltonian and grid. The real part of the autocorrelation is not shown here because the wavefunction correction does not alter $E_\text{dynamic}$ and therefore leaves the oscillation period of $\mathrm{Re}(A(t))$ unchanged.

It should be emphasised that if the Hamiltonian itself suffers from discretisation errors, either the initial wavefunction correction or ITP will faithfully return the ground state of that imperfect operator. These techniques only improve the stability of the time evolution, while more accurate dynamical energies require prior refinement of the discretised Hamiltonian, such as through the potential operator correction.

\begin{figure}[!htbp]
    \centering
    \includegraphics[scale=0.15]{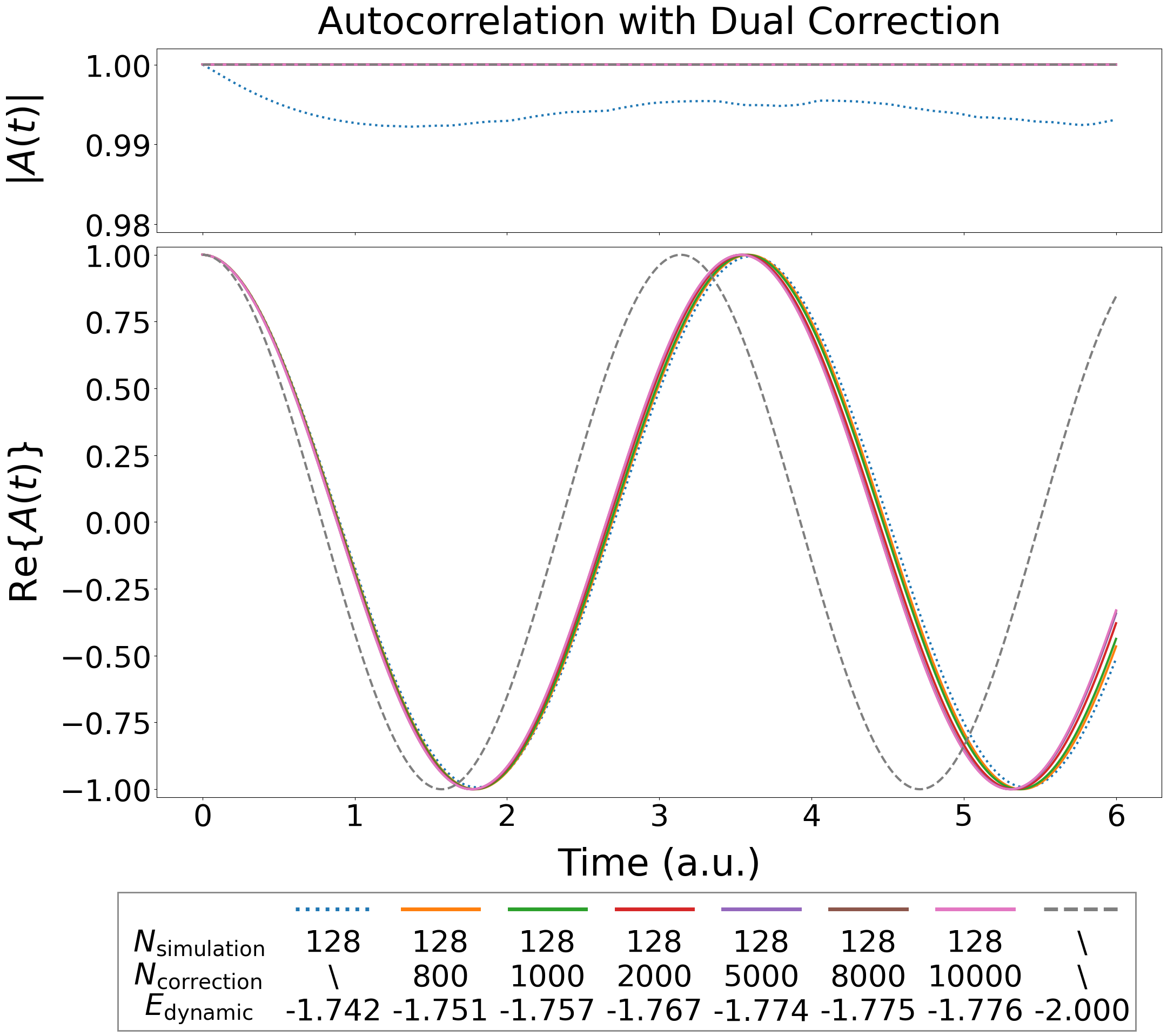}
    \caption{\justifying Absolute value and real part of autocorrelation functions recorded from simulations of the 2D hydrogen using $V_\text{corrected}$ computed from different $N_\text{correction}$ and corrected wavefunctions. Analytical profiles and results from uncorrected runs are plotted with dashed and dotted linestyles, respectively.}
    \label{dual corr}
\end{figure}

While the potential correction optimises $E_\text{dynamic}$ and the wavefunction correction preserves time fidelity, we have examined that they together produce synergistic improvements.
In simulations of the 2D hydrogen system that simultaneously apply both corrections, the absolute autocorrelation remains ideal throughout, and the numerical energy $E_\text{dynamic}$ fully exhibits the benefit of the corrected potential as shown in Figure~\ref{dual corr}. This dual-correction strategy is particularly suited for systems where increasing $N_{\text{simulation}}$ is computationally expensive, as it enables better-quality simulations than uncorrected baselines when holding the grid density unchanged.


\section{Quantum Implementation of Corrected Coulombic System Dynamics}
\label{QuanMap}


With the rapid progress in quantum computing, simulations of real-space Coulombic systems are expected to become significantly more tractable and efficient once practical quantum simulators are available~\cite{low2025}. We therefore seek to investigate the efficient realisation of our algorithm on quantum computers.

\begin{figure}[!htbp]
\centering
\scalebox{1.1}{
\Qcircuit @C=1em @R=1.5em {
\lstick{\text{Ancilla} \ket{0}} & \gate{H} & \ctrl{1} & \gate{H}  & \meter  \\
\lstick{\text{System} \ket{\psi}} & \qw {/}^{n} & \gate{U} & \qw & \qw  
}}
\caption{\justifying Schematic circuit for measuring the real part of autocorrelation functions at a specific time step.}
\label{had test}
\end{figure}
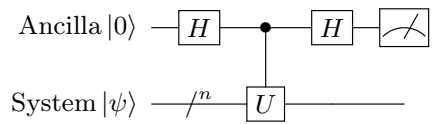

\begin{figure*}[!htbp]
    \centering
    \includegraphics[scale=0.13]{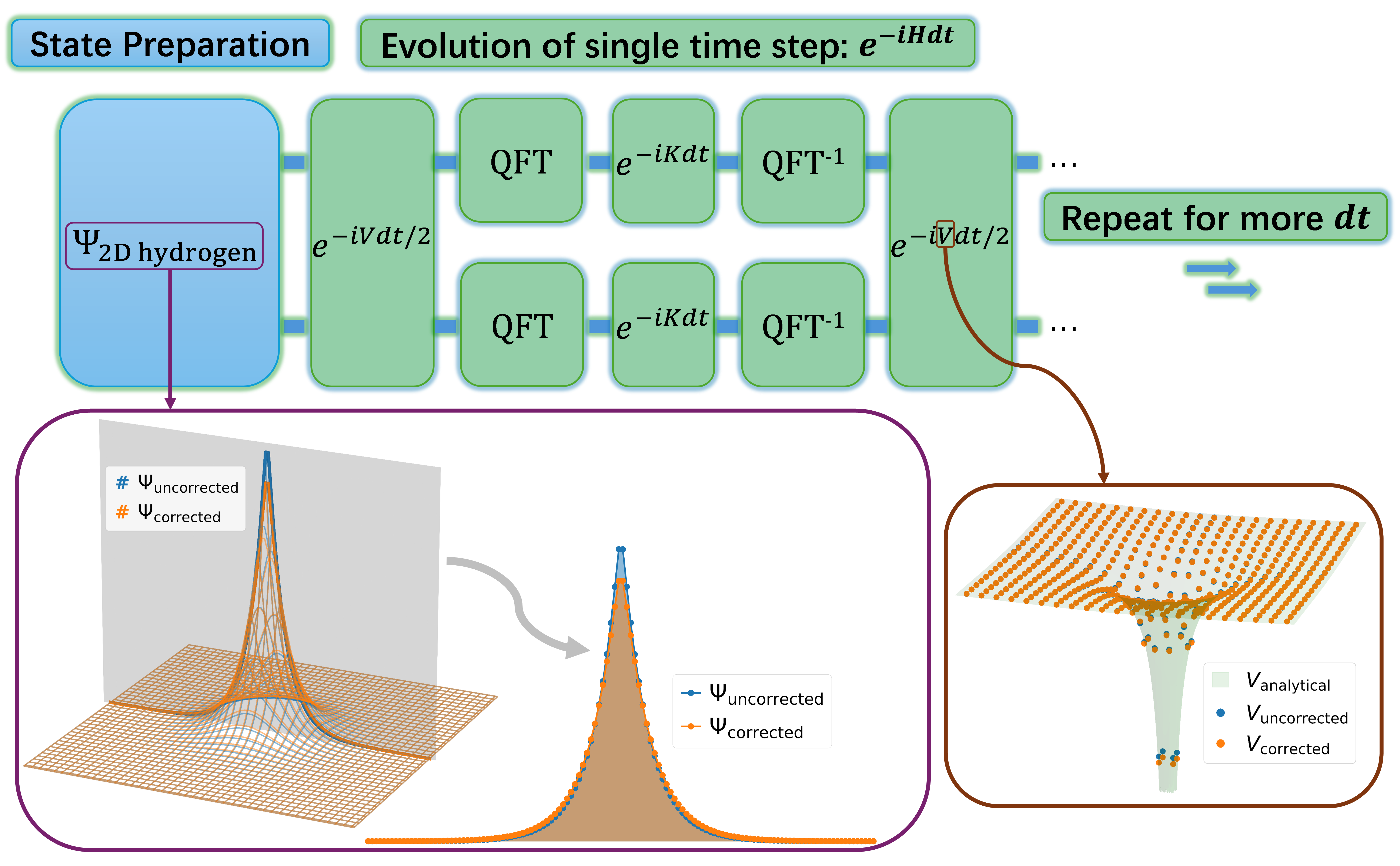}
    \caption{\justifying Schematic overview of essential components for simulating the 2D hydrogen system.}
    \label{timevofig}
\end{figure*}


The key properties of interest in this work, $E_\text{dynamic}$ and time fidelity, are both derived from the time-dependent autocorrelation function. On quantum platforms, the extraction of autocorrelation values corresponds to the measurement stage of the relevant quantum algorithms and is typically realised through the single-ancilla Hadamard test. The Hadamard test is an interferometric technique widely adopted for extracting inner products and expectation values~\cite{Patel2025,Ding2023, Mitarai2019, Abhijith2022, Aharonov2009}. 
As shown in Figure~\ref{had test}, it works by preparing an ancillary qubit in a superposition state, applying controlled time evolution conditioned on this qubit up to a given time step, and computing the autocorrelation from measurement statistics of the ancilla. 
Repeated measurements yield the probabilities of observing $\ket{0}$ and $\ket{1}$, from which the real part of the autocorrelation at this time step is computed as 
$$\operatorname{\mathbb{R}e}\{A(t)\}=P_{\ket{\text{ancilla}}}(0)-P_{\ket{\text{ancilla}}}(1).$$
Similarly, the imaginary part of the autocorrelation is obtained via 
$$\operatorname{\mathbb{I}m}\{A(t)\}=P_{\ket{\text{ancilla}}}(1)-P_{\ket{\text{ancilla}}}(0),$$
by inserting an additional $\mathbf{S}$ gate before the final Hadamard gate.

The measurement count serves as one of the metrics when assessing quantum resources. This cost can be reduced by employing a sparser sampling strategy, i.e., measuring the ancilla at selected time points with wider intervals, provided the resulting autocorrelation remains sufficiently smooth to resolve the desired properties.

Another essential factor in resource evaluation is the circuit depth, which depends on the specific structure of the quantum circuits. In addition to autocorrelation measurements, the complete circuit also involves initial state preparation and time propagation stages. 
For these two stages, we have proposed and analysed two targeted correction schemes in previous sections, both of which are compatible with quantum architectures. 
In the remainder of this section, we focus on the 2D Hydrogen system as an illustrative example to detail the application of these correction schemes on realistic quantum hardware, along with an analysis of the associated gate depths and resource implications for constructing the circuit  shown in Figure~\ref{timevofig}.

\subsection{Preparation of Corrected Wavefunction}

This section describes the quantum circuit construction for encoding the corrected wavefunction onto a register that represents a two-dimensional spatial space.
We allocate $n=\log_2{N_\text{simulation}}$ qubits per spatial dimension to represent a 2D $N_\text{simulation} \times N_\text{simulation}$ grid. The resulting quantum register consists of two $n$-qubit sub-registers, initialized in the state $\ket{0}^{\otimes 2n}$. The goal is to transform this initial state into a superposition state that reflects the amplitudes of the corrected wavefunction, i.e.,
$$\sum^{2^n-1}_{j_1=0}\sum^{2^n-1}_{j_2=0} c_{j_1j_2}\ket{j_1}\otimes \ket{j_2},$$
where the coefficients $c_{j_1j_2}$ denote the discretised amplitudes. Each basis state $\ket{j_1} \otimes \ket{j_2}$ corresponds to a specific grid point within the 2D space.

One straightforward approach for preparing this quantum state is to treat the two spatial dimensions as a single index and load the $2^{2n}$ amplitudes directly onto a $2n$-qubit register. This can be implemented by a cascade of uniformly-controlled rotations~\cite{Mttnen2004} requiring a gate count of $2^{2n+1}-3$, where only two-qubit $\mathbf{CNOT}$ gates and single-qubit $\mathbf{R_y}$ gates are used due to the real and non-negative nature of the amplitudes. For $n=\log_2{128}=7$, the total gate count is estimated to be 32$,$765.


To reduce the gate depth and improve efficiency, we alternatively adopt the Fourier Series Loader (FSL) method~\cite{moosa2023}, which constructs the multidimensional quantum state from truncated Fourier series representations. Instead of loading all $2^{2n}$ amplitudes directly, only $2^{2l}$ Fourier coefficients with $l < n$ are used to approximate the full wavefunction.


\begin{figure}[!htbp]
    \centering
    \includegraphics[scale=0.15]{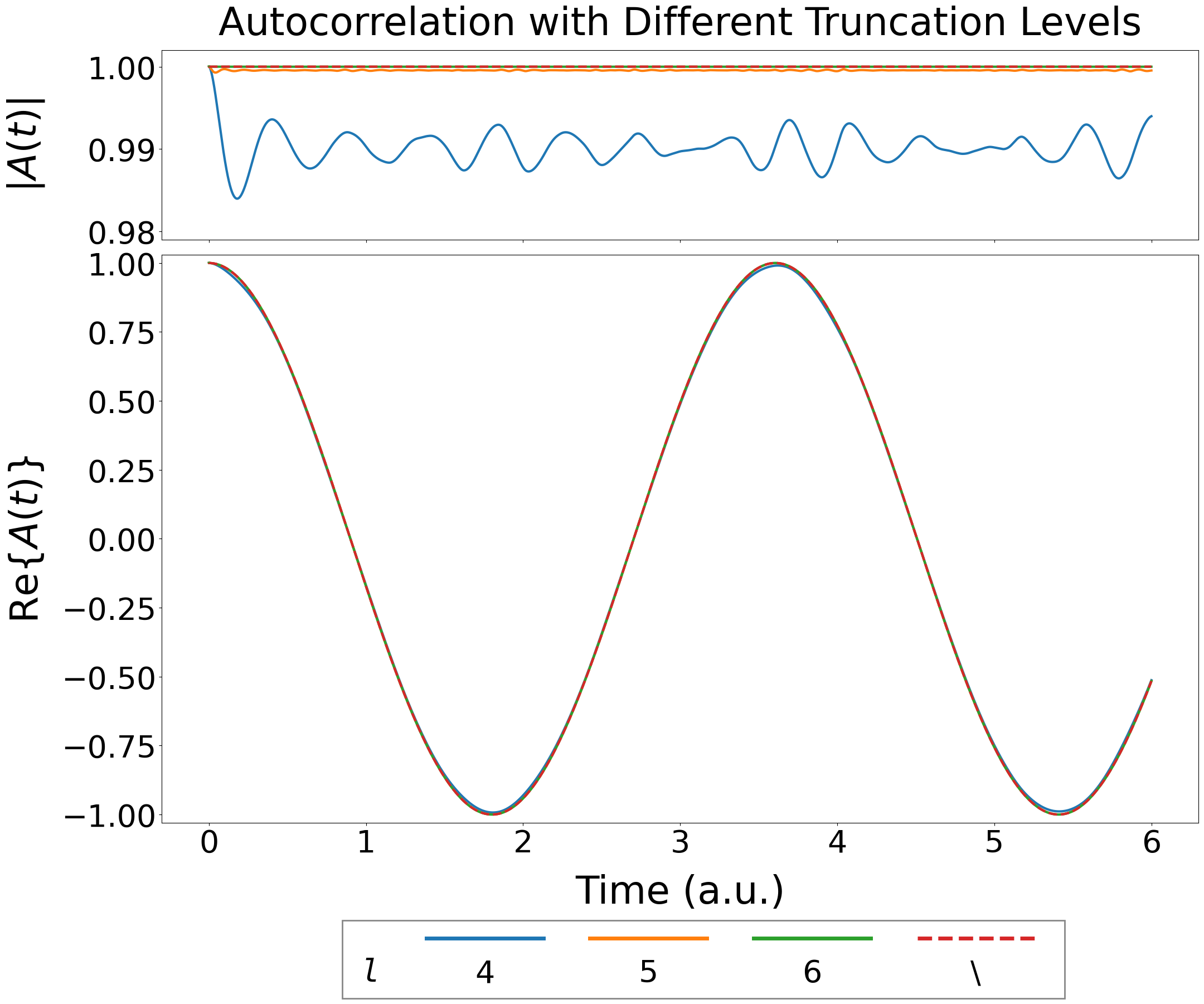}
    \caption{\justifying Autocorrelation functions recorded from 2D hydrogen simulations initialized with corrected wavefunctions prepared by FSL method. $l$ represents the truncation level used in the state preparation. Benchmark data simulated from the exact corrected wavefunction ($n=7$) are displayed in dashed linestyle.}
    \label{corrwf prep}
\end{figure}

\begin{figure*}[!htbp]
\centering
\scalebox{1.2}{
\Qcircuit @C=1em @R=1.5em {
\lstick{\ket{k_{1}}} & \ctrl{3} &\qw &\qw &\ctrl{3} & \qw& \qw & \ctrl{3} &\qw &\qw &\ctrl{3} & \qw& \qw& \qw\\
\lstick{\ket{k_{2}}} & \qw &\qw &\qw &\qw & \ctrl{2} & \qw & \qw &\qw &\qw &\qw & \ctrl{2} & \qw& \qw\\
\lstick{\ket{k_{3}}} &\qw  & \ctrl{1}& \qw & \qw & \qw & \qw &\qw  & \ctrl{1}& \qw & \qw & \qw & \qw& \qw\\
\lstick{\ket{k_{4}}} & \targ & \targ & \gate{R_{z}(2a_{12})} & \targ & \targ & \gate{R_{z}(2a_{15})}& \targ & \targ & \gate{R_{z}(2a_{10})} & \targ & \targ & \gate{R_{z}(2a_{9})} & \qw
}}
\caption{\justifying Circuit fragment corresponding to the $b=4$ partition group. The four remaining indices $\{12,15,10,9\}$ are arranged in gray-code order.}
\label{correop circ}
\end{figure*}
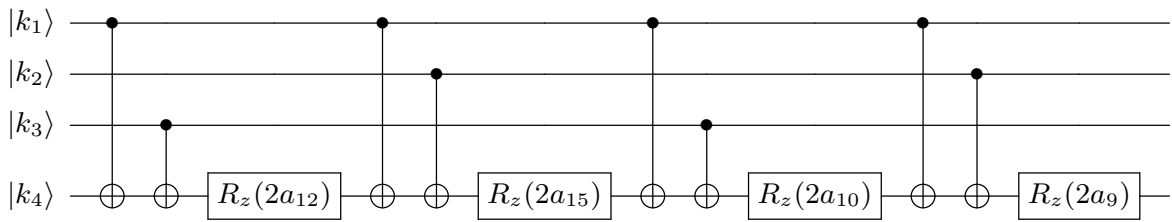

A truncated set of Fourier coefficients is first precomputed classically from the corrected wavefunction and used to initialize a compact quantum state onto a $2l$-qubit register. The loading circuit implements sequences of uniformly controlled rotations including alternating layers of $\mathbf{CNOT}$ gates interleaved with both $\mathbf{R_y}$ and $\mathbf{R_z}$ gates that result in total $2(2^{2l+1}-3)$ gates. 
The remaining $n-l$ qubits per dimension are then activated via entangling operations of $n-l$ $\mathbf{CNOT}$ gates, which effectively expands the state to the full $2n$-qubit register by zero-padding. Finally, an inverse QFT maps the encoded amplitudes to real-space. The total number of gates for this FSL procedure is denoted by $2^{2l+2}+n^2/2+2n-l-13/2$ (for odd $n$).


The choice of truncation level $l$ depends on the trade-off between simulation fidelity and gate complexity. For instance, by fixing $n = 7$, we benchmark FSL-based initial states prepared from Qiskit~\cite{Qiskit,moosa2023} with varying $l$ values against the exact corrected wavefunction. Figure~\ref{corrwf prep} shows that with $l=6$, the time-dependent autocorrelation remains fully consistent with the exact result. The corresponding gate count, 16$,$410, is nearly half of that required by full-amplitude loading. Further reduction to $l=5$ introduces minor deviations and undesirable oscillations in $A|(t)|$, but still fundamentally preserves the global stability of autocorrelation behaviours.




\subsection{Implementation of Time Propagation} 
\label{WHT}

Simulating quantum dynamics on digital quantum hardware requires careful construction of the time evolution operator using gate-efficient techniques. 
In this section, we present the quantum implementation of time propagation under the Trotterized Hamiltonian. The full circuit is composed of three fundamental components: the kinetic evolution operator, the potential evolution operator, and the QFT that facilitates conversion between position and momentum spaces, as illustrated in Figure~\ref{timevofig}. Here the QFT and its inverse block are applied before and after the kinetic term, respectively, each introducing $n^2/2+n-1/2$ gates for odd $n$.

The kinetic evolution operator takes the form of an exponentiated quadratic polynomial in the momentum basis, which can be succinctly complied into successive controlled $\mathbf{U_{1}}$ gates~\cite{Pauline2020}. For the 2D hydrogen system, it factorizes along dimensions and can be applied in parallel across the respective $n$-qubit sub-registers:
$$e^{-iKdt} =e^{(\frac{-ip_x^{2}dt}{2})}e^{(\frac{-ip_y^{2}dt}{2})},$$
resulting in a gate count of $n^2$ per Trotter step. 

We then turn to the potential evolution operator, 
$e^{-iV_\text{corrected}dt/2}$,
which is implemented directly in the position basis.  
A compact and ancilla-free method for realising such diagonal unitary by Walsh-operator formalism has been proposed by~\cite{Welch2014}, which first maps the target unitary into a product of exponentials of Walsh Operators, and then reorder them to reduce repeated $\mathbf{CNOT}$ gates. 

We begin with computing the Walsh–Hadamard coefficients $a_j$s via the Walsh-Hadamard Transform (WHT) of the discrete $F=-V_\text{corrected}dt$ (including $N_\text{total}=2^{2n}$ real values). 
The potential evolution operator $e^{iF}=e^{-iV_\text{corrected}dt}$ therefore maps to the quantum gate sequence $$\prod_{j=0}^{N_\text{total}-1} e^{-ia_j\omega_j},$$ where $\omega_j$ is the Walsh operator indexed by the Paley order $j$.

These $j$ values are grouped by the most significant non-zero bit, $b$, of their binary representation $\ket{j}$. 
Within each such partition, the circuit implements alternating layers of $\mathbf{R_z}(2a_j)_{[b]}$ and $\mathbf{CNOT}_{[b]}^{[c]}$ gates, with the subscript and superscript representing target and control qubits, respectively. 
Specifically, a $\mathbf{CNOT}_{[b]}^{[c]}$ gate is introduced only when the $c$th bit of the bitwise XOR between adjacent $\ket{j}$ is non-zero. Additional $\mathbf{CNOT}_{[b]}^{[c]}$ gates accounting for the XOR between the first and last $\ket{j}$ in each group are inserted at its beginning.

As shown in~\cite{Welch2014}, an optimal circuit construction with $2^{2n+1}-3$ gates is achieved by ordering the $N_\text{total}$ $\mathbf{R_z}(2a_j)_{[b]}$ gates in Gray‑code sequence, which leaves just a single $\mathbf{CNOT}$ between successive $\ket{j}$ values.
In practice, however, the number of non-trivial Walsh components is often much smaller than $N_\text{total}$, and discarding those $a_j$ near zero significantly simplifies the circuit while preserving autocorrelation behaviours.

For example, with $n=7$ qubits per sub-register, we find that retaining only the largest $N_\text{approx} = 2^{12}$ coefficients (instead of the full $2^{14}$) still generates $E_\text{dynamic}$ and time fidelity indistinguishable from using the exhaustive series. Consequently, the number of $\mathbf{R_z}(2a_j)$ gates reduces to $2^{12}=4096$. 
Arranging this truncated set of $2^{12}$ indices in Gray‑code order produces 8312 $\mathbf{CNOT}$ gates. Therefore, the total gate count for implementing $e^{-iV_\text{corrected}dt/2}$ is 12408. Figure~\ref{correop circ} illustrates a representative fragment of the constructed circuit corresponding to the partition group of $b=4$, where only four $j$ values, $\{12,15,10,9\}$, survive the truncation.

Following the structure shown in Figure~\ref{timevofig}, the propagation of one single time step $dt$ includes the full‑step kinetic evolution, both half‑step potential evolutions, and double QFTs, which amounts to 24927 gates in total. For 6000 steps, the circuit depth scales linearly to $\sim 1.5\times10^{8}$ gates.



\section{Conclusions}
The grid-based SO-FT method employed in this work has wide applicability in simulating time-propagated quantum dynamics. However, its accuracy degrades in the presence of Coulombic singularities, as the associated infinite value cannot be adequately represented using finite grids. While increasing grid density mitigates this error by more closely depicting the divergence, the exponential scaling of computational memory presents a major burden, especially for simulations with numerous Trotter steps. We therefore develop two correction schemes, one targeting Hamiltonian and the other focusing on wavefunction, to improve the robustness of Coulombic system simulations while maintaining the existing coarse grids.

The first correction reformulates the Coulomb potential by calculating its expectation value under the actual grid basis functions. This correction is pre-computed once and does not inflate costs during the time evolution. Our simulations in 1D electron-electron quantum ring and 2D nucleus-electron hydrogen systems validate the ability of the corrected potential to reduce bias in numerical energy and stabilize time fidelity. The second scheme concerns a wavefunction correction based on analytic solutions of soft Coulomb potentials. By incorporating additional polynomials and energy terms calibrated to specific grid resolution into the initial state, we create an initial state for 2D hydrogen that better represents the eigenstate of its grid-based Hamiltonian. Starting from this corrected wavefunction yields near-perfect time fidelity throughout the evolution window.

Although the corrections do not reproduce the exact singular behaviour, they consistently deliver smaller errors than the uncorrected forms. Because the associated deviation is systematic rather than stochastic, e.g., manifesting as quantifiable differences between corrected and analytical ground-state energies, it can be readily incorporated into analyses and practically subtracted as a correctible offset if necessary. Our correction schemes may offer tangible benefits when applied to more complex scenarios, such as simulations of multi-component systems involving hydrogen interacting with heavier atoms or under the influence of spatially varying external potentials (e.g., electric or magnetic).
In such settings, small numerical inaccuracies can be amplified and distort long-range interactions.
The outlined strategies help reduce cumulative numerical errors and facilitate improved stability, leading to more reliable predictions of system dynamics. 

Extending the classical grid-based SO-FT method to its quantum analogue is straightforward, owing to the structural compatibility of grid-based representations with quantum registers and the inherent efficiency of QFT implementations. 
Within this framework, the correction schemes proposed in this work also adapt readily to quantum computing architectures. The corrected potential operator can be implemented using a truncated Walsh expansion, while the corrected wavefunction is prepared through a truncated Fourier series.
By appropriately selecting truncation thresholds to preserve simulation accuracy, we estimate a circuit depth of $\sim 1.5\times10^{8}$ gates for simulating a 2D hydrogen system with $n=7$ qubits per dimension over 6,000 Trotter steps.

Further optimisations would broaden the scope and efficiency of these corrections. 
On the algorithmic side, incorporating off-diagonal contributions into the potential correction may improve accuracy by accounting for inter-basis overlaps that are currently neglected. Similarly, generalising the wavefunction correction to more Coulombic systems will expand its versatility. In this study, we have chosen model systems free from antisymmetrization requirements to address grid discretisation errors that are independent of particle exchange symmetry. Since the potential correction only involves the operator, not the wavefunction, we anticipate that it will also work well in multi-electron systems with proper antisymmetrization. From a quantum computing perspective, these correction schemes are compelling for minimising error accumulation without demanding additional qubits, which is particularly advantageous in resource-constrained regimes. Embedding them into large-scale models could enable high-fidelity quantum simulations under realistic computational budgets.

\section{Author Contributions}
Xiaoning Feng conceived the project, developed the algorithm, performed the simulations, carried out the data analysis, and wrote the original draft of the manuscript. David P. Tew supervised the research and provided scientific guidance on the research design and interpretation of the results. Hans Hon Sang Chan provided supportive feedback through discussions. 


\section{Conflicts of Interest}
The authors declare that they have no competing interests.

\bibliographystyle{apsrev4-1.bst}
\bibliography{references.bib}

\end{document}